\documentclass[floatfix,showpacs,preprintnumbers,amsmath,amssymb]{revtex4}

\usepackage{bm}
\usepackage{latexsym}
\usepackage{eufrak}
\usepackage[latin1]{inputenc}
\usepackage{latexsym}
\usepackage{amsmath}
\usepackage{epsfig}
\usepackage{ifpdf}
\usepackage{graphicx}
\usepackage{dcolumn}

\def\be{\begin{equation}}
\def\ee{\end{equation}}
\def\bea{\begin{eqnarray}}
\def\eea{\end{eqnarray}}
\def\real{\hbox{{I}\kern-.2em{\bf R}}}

\begin{document}

\title{Class of Einstein-Maxwell Phantom Fields:
\\Rotating and Magnetised Wormholes}
\author{Tonatiuh Matos\footnote{Part of the Instituto Avanzado de Cosmolog\'ia (IAC) collaboration http://www.iac.edu.mx/}} 
\email{tmatos@fis.cinvestav.mx}
\affiliation{Departamento de F{\'\i}sica, \\
Centro de Investigaci{\'o}n y Estudios Avanzados del I. P. N.,\\
A. P. 14-700, 07000 M{\'e}xico, D.F.,M\'exico}

\begin{abstract}
Using a new ansatz for solving the Einstein equations with a scalar field with the sign of the kinetic term inverted, I find a series of formulae to derive axial symmetric
stationary exact solutions of the Phantom scalar field in general relativity. We focus on the solutions which represent wormholes. The
procedure presented in this work allows to derive new exact solutions up to
very simple integrations. Among other results, I find exact rotating
solutions containing magnetic monopoles, dipoles, etc., coupled
to Phantom scalar and to gravitational multipole fields.
\end{abstract}

\draft


\date{\today}
\maketitle

\section{Introduction}
Doubtless we are living exiting times, 10 years ago astronomers discovered an invisible component of the universe which represents more than 70\% of the total matter of the universe and provokes that the universe expands accelerating. This mysterious component dominates the universe and avoid that the galaxies clusters collapse too much. There are several candidates for the nature of this dark energy. The most accepted candidate is the cosmological constant, but other candidates lake a scalar field (Quintessence) or k-essence (see for example \cite{quintessence} and \cite{kessence} and references derin), among others, are intensively studied. One of the most exotic proposals to be the dark energy is the Phantom field, a scalar field with the kinetic term with opposite sign. This Phantom energy violates the energy conditions and contains an exotic thermodynamics. Nevertheless, nobody knows the nature of the dark energy, thus nobody knows if this dominating component of the universe should fulfil the standard physics. In principle this new component could be something which contains new physics, new thermodynamics and new energy conditions. On the other side, the violations of the energy conditions in the way as phantom energy do, can be a source of wormholes \cite{EO}. The existence of wormholes in the universe is very interesting because they could be highways to visit stars and galaxies, otherwise it will be impossible to go enough far away to visit other worlds. Of course this fact is very speculative to this stage, but no observation can discard the phantom energy as the dominant component in the universe \cite{sahni} and the fact that phantom energy can be the source of wormholes \cite{ER}-\cite{EO} is more than exciting to investigate about the existence of stars made of this kind of matter.

In this work I pretend to give a method to study the Einstein equations with phantom energy as source. One of the main problems of Einstein-phantom solutions is that all the well know ones are unstable \cite{shin}, \cite{paco1}. In \cite{dario} 
we conjectured that rotation or the magnetic field could stabilise a phantom star. Here I give a general method, based in old generation methods of exact solutions of the Einstein field equations \cite{DoctorN}-\cite{manuque}, to obtain exact solutions of the Einstein equations with phantom matter source. 

Thus I start with the Lagrangian (compare for example with \cite{maridari}, \cite{gabino})
\begin{equation}
{\cal L}=\sqrt{-g}\,(-R-2\,(\nabla \,\phi )^{2}+e^{-2\,\alpha \,\phi
}\,F^{2}),  \label{lag0}
\end{equation}
where $g$ is the determinant of the metric tensor, $R$ is the scalar
curvature, $\phi $ the Phantom field, $F$ the Maxwell one and $\nabla$ is the covariant derivative. The constant $\alpha $ is a free parameter which governs the strength of the coupling of
the Phantom to the Maxwell field. When $\alpha =0$, the action reduces to
the uncoupled Einstein-Maxwell phantom theory. When $\alpha \neq1$, the action is a theory of phantom field with a dilaton like coupling with electromagnetism. We will consider this theory for all values of $\alpha\neq0 $.

The field equations derived from Lagrangian (\ref{lag0}) are give by (see also \cite{maridari})
\begin{eqnarray}
\nabla _{\mu }(e^{-2\alpha \phi }F^{\mu \nu }) &=&0;  \nonumber \\
\nabla ^{2}\phi -{\frac{\alpha }{{2}}}e^{-2\alpha \phi }F^{2} &=&0; 
\nonumber \\
-2\nabla _{\mu }\nabla _{\nu }\phi +2e^{-2\alpha \phi
}(F_{\mu \rho }{F_{\nu }}^{\rho }-{\frac{1}{{2}}}g_{\mu \nu }e^{-2\alpha
\phi }F^{2})&=&R_{\mu \nu } .  \label{eqL}
\end{eqnarray}

In this work I am interested in isolated phantom stars, $i.e.$, in stars made of Phantom matter and an electromagnetic field. We can suppose that these stars are axial symmetric and stationary.
Thus, I will analyse spacetimes characterised by two Killing vector fields $X$
and $Y$ and introduce coordinates $t$ and $\varphi $ which are chosen such
that $X={\frac{{\partial }}{{\partial \,t}}}$ and 
$Y={\frac{{\partial }}{{\partial \,\varphi }}}$. The corresponding line element can then be expressed as \cite{kra}
\begin{equation}
ds^{2}=-f\,(dt-\omega\,d\varphi )^{2}+f^{-1}[e^{2k}(d\rho ^{2}+dz^{2})+\rho
^{2}d\varphi ^{2}],  \label{papa}
\end{equation}
where $f,\,\omega $, and $k$ are functions of $\rho $ and $z$ only. The
electromagnetic potential has the form $A_{\mu }=(A_{0},0,0,A_{3})$, and
again $A_{0},A_{3}$, and the Phantom field, $\phi $ are functions of $\rho 
$ and $z$ only.
In Boyer-Lindquist coordinates $\rho=\sqrt{r^2-2\,m\,r+\sigma^2}\,\sin(\theta),\,z=(r-m)\,\cos(\theta)$, space-time metric (\ref{papa}) reads
\begin{eqnarray}
ds^{2}&=&-f\,(dt-\omega\,d\varphi)^{2}+{f}^{-1}\,\left[ K\,dr ^{2}+(r^2-2\,m\,r+\sigma^2)\,\left(K\, d\theta^{2}+\sin^2(\theta)\,d\varphi ^{2}\right)\right]  ,  \label{papa_BL}
\end{eqnarray}
where 
\[
 K=\frac{(r-m)^2+(\sigma^2-m^2)\,\cos^2(\theta)}{(r^2-2\,m\,r+\sigma^2)}\,e^{2k}
\]

The main goal of this work is to give solutions of the Einstein field equations using metric (\ref{papa_BL}).

\section{Functional Space Formulation}

In what follows I will introduce the functional geodesic formulation for
Lagrangian (\ref{lag0}) (compare with \cite{maridari} and \cite{gabino}), the method is fully explained in \cite{DHA}.
Essentially, the formulation takes advantage of the fact that the
introduction of a line element in the Lagrangian for the Einstein-Hilbert
action (\ref{lag0}), and performing the variation, there are operations which
commute for some cases, in particular for the axisymmetric stationary one.
Thus, introducing the operator $D=(\partial _{\rho },\partial _{z})$, taking
out a total divergence term and eliminating the terms with $D\,k$ by means
of a Legendre transformation, I obtain that the original Lagrangian, given
by (\ref{lag0}), can be rewritten as 
\begin{equation}
{\cal L}={\frac{\rho }{{2\,f^{2}}}}\,Df^{2}-{\frac{{f^{2}}}{{2\,\rho }}}
\,D\omega ^{2}-{\frac{{2\,\rho }}{{\alpha ^{2}\,\kappa ^{2}}}}\,D\,\kappa
^{2}+{\frac{{2\,f\,\kappa ^{2}}}{{\rho }}}\,[(\omega \,DA_{0}+DA_{3})^{2}-{%
\frac{{\rho ^{2}}}{{f^{2}}}}\,DA_{0}^{2}],  \label{lag1}
\end{equation}
where $\kappa ^{2}=e^{-2\,\alpha \,\phi }$. Here I will take only space-time with $\alpha\neq0$.
The Euler-Lagrange equations, obtained directly from extremizing the action
for such Lagrangian $D({\frac{{\partial {\cal L}}}{{\partial D\,Z^{a}}}})-({%
\frac{{\partial {\cal L}}}{{\partial Z^{a}}}})=0$, with $Z^{a}=(f,\omega
,A_{0},A_{3},\kappa )$, are
\begin{subequations}
\begin{eqnarray}
\text{The Klein-Gordon equation:}\hskip5cm&&\nonumber\\
 D^{2}\,\kappa +({\frac{{D\,\rho }}{\rho }}-{\frac{2\,{D\,\kappa }}{\kappa }}
)\,D\kappa +{\frac{{\alpha ^{2}\,\kappa ^{3}\,f}}{{\rho ^{2}}}}[(\omega
\,DA_{0}+DA_{3})^{2}-{\frac{{\rho ^{2}}}{{f^{2}}}}\,DA_{0}^{2}]&=&0,\\
\label{eq:Ein_Klein-Gordon}
\text{The Maxwell Equations:}\hskip5cm\nonumber&& \\
D\left( {\frac{{f\,\kappa ^{2}}}{{\rho }}}\,(\omega \,DA_{0}+DA_{3})\right)
&=&0,  \nonumber \\
D\left( \kappa ^{2}\,[{\frac{{f\,\omega }}{{\rho }}}(\omega \,DA_{0}+DA_{3})-%
{\frac{{\rho }}{{f}}}\,DA_{0}]\right) &=&0,  \label{eq:Ein_maxw}\\
\text{The main Einstein's Equations:}\hskip5cm\nonumber&&\\
D^{2}\,f+({\frac{{D\,\rho }}{\rho }}-{\frac{{D\,f}}{{f}}})\,D\,f+
{\frac{f^{3}}{\rho ^{2}}}\,D\,\omega ^{2}-{\frac{{2\,\kappa ^{2}\,f^{2}}}{{\rho ^{2}
}}}\,[(\omega \,DA_{0}+DA_{3})^{2}+{\frac{{\rho ^{2}}}{{f^{2}}}}
\,DA_{0}^{2}]&=&0,   \label{eq:Ein_f}\\
D^{2}\,\omega -({\frac{{D\,\rho }}{\rho }}-{\frac{{2\,D\,f}}{{f}}}
)\,D\,\omega +{\frac{{4\,\kappa ^{2}}}{{f}}}\,(\omega
\,DA_{0}+DA_{3})\,D\,A_{0}&=&0,  \label{eq:Ein_w}
\end{eqnarray}
\label{eq:Ein}
\end{subequations}
Observe that the metric
function $k$ does not appear in the field equations (\ref{eq:Ein}), it is determined by
quadratures in terms of the rest of the functions. The Lagrangian (\ref{lag1}) sometimes is viewed as describing the line element in the potential space \cite{DoctorN},
thus the motion equations can be seen as geodesics in this potential space.

We define the differential operator ${\tilde{D}}=(-\partial _{z},\partial _{\rho })$, from this definition it follows that $D\,{\tilde{D}}=0$ for any analytic function. Thus we conclude from the second
Maxwell equation (\ref{eq:Ein_maxw}), that there exits a potential \cite{DoctorN} $\chi $,
such that 
\begin{equation}
{\tilde{D}}\,\chi ={\frac{{2\,f\,\kappa ^{2}}}{\rho }}\,(\omega
\,DA_{0}+DA_{3}).  \label{chi}
\end{equation}
And with this potential, the second Einstein's equation (\ref{eq:Ein_w}), can be rewritten as $D({\frac{{f^{2}}}{\rho }}\,D\,\omega +\psi \,{\tilde{%
D}}\,\chi )=0$, with $\psi =2\,A_{0}$, so that there exists another
potential \cite{DoctorN} $\epsilon $, defined by 
\begin{equation}
{\tilde{D}}\,\epsilon ={\frac{{f^{2}}}{\rho }}\,D\,\omega +\psi \,{\tilde{D}}%
\,\chi .  \label{eps}
\end{equation}

The use of these potentials $\chi $ and $\epsilon $, will be helpful in the
procedure of defining harmonic functions, so I rewrite the field equations
in terms of the functions $(f,\epsilon ,\chi
,\psi ,\kappa )$ as
\begin{subequations}
 \begin{eqnarray}
D(\,\rho \,D\,\kappa) -\rho\,{\frac{2\,{D\,\kappa }\,D\kappa}{\kappa }} -{\frac{{\rho\,\alpha ^{2}\,\kappa ^{3}}}{{4\,f}}}(D\psi ^{2}-{\frac{1}{%
{\kappa ^{4}}}}\,D\,\chi ^{2}) &=&0,  \label{eq:Pot_kappa} \\
D(\rho\,D\,\psi) +\rho\,\left( {\frac{{2\,D\,\kappa }}{\kappa }}-{%
\frac{{D\,f}}{{f}}}\right) \,D\,\psi -{\frac{\rho}{{\kappa ^{2}\,f}}}\,(D\,\epsilon
-\psi \,D\,\chi )\,D\,\chi &=&0,  \label{eq:Pot_psi} \\
D(\rho\,D\,\chi) -\rho\,\left( {\frac{{2\,D\,\kappa }}{\kappa }}+{\frac{{D\,f}}{{f}}}\right) \,D\,\chi +{\frac{{\rho\,\kappa ^{2}}}{{f}}}\,(D\,\epsilon
-\psi \,D\,\chi )\,D\,\psi &=&0,  \label{eq:Pot_chi} \\
D(\rho\,D\,f)-\rho\,{\frac{{D\,f}\,D\,f}{{f}}}+{\frac{\rho}{{f%
}}}(D\,\epsilon -\psi \,D\,\chi )^{2}-{\frac{{\rho\,\kappa ^{2}}}{{2}}}\,(D\psi
^{2}+{\frac{1}{{\kappa ^{4}}}}\,D\,\chi ^{2}) &=&0,  \label{eq:Pot_f} \\
D(\rho\,(D\,\epsilon -\psi \,D\chi))-\rho\,D\,\psi \,D\,\chi  -{\frac{{2\rho\,D\,f}}{{f}}}\,(D\,\epsilon -\psi \,D\,\chi ) &=&0,
\label{eq:Pot_epsilon}
\end{eqnarray}
\label{eq:Pot}
\end{subequations}
where I have used the fact that ${\tilde{D}}A\,{\tilde{D}}B=DA\,DB$, for
any functions $A,B$. The equation for $\chi $ is obtained from $D\,{\tilde{D}%
}\,A_{3}=0$, and the one for $\epsilon $ is obtained from $D\,{\tilde{D}}%
\,\omega =0$. 

The remaining metric function $k$ is determined by
quadratures in terms of the other field functions \cite{manuque}, explicitly I have 
\begin{eqnarray}
k_{,\rho } &=&{\frac{\rho }{{4\,f^{2}}}}[{f_{\rho }}^{2}-{f_{z}}^{2}+{
\epsilon _{\rho }}^{2}-{\epsilon _{z}}^{2}+(\psi^2-{\frac{f}{\kappa ^{2}}}%
)\, ({\chi _{\rho }}^{2}-{\chi _{z}}^{2})-2\,\psi (\epsilon _{\rho }\,\chi
_{\rho }-\epsilon _{z}\,\chi _{z}) \nonumber\\
&&-f\,\kappa ^{2}({\psi _{\rho }}^{2}-{\psi _{z}}^{2})-({\frac{{2\,f}}{{%
\alpha \,\kappa }}})^{2}\,({\kappa _{\rho }}^{2}-{\kappa _{z}}^{2})], \nonumber\\
k_{,z} &=&{\frac{\rho }{{2\,f^{2}}}}[f_{\rho }\,f_{z}+\epsilon _{\rho
}\,\epsilon _{z}-\kappa ^{2}\,f\,\psi _{\rho }\,\psi _{z}-\psi \,(\epsilon
_{\rho }\,\chi _{z}+\epsilon _{z}\,\chi _{\rho })+ \nonumber\\
&&(\psi^2-{\frac{f}{\kappa ^{2}}})\,\chi _{\rho }\,\chi _{z}-({\frac{{2\,f}}{%
{\alpha \,\kappa }}})^{2}\,\kappa _{\rho }\,\kappa _{z}].  
\end{eqnarray}
or using the complex variable $\zeta=\rho+i\,z$ I obtain that
\begin{eqnarray}
 k_{,\zeta} &=&{\frac{\rho }{{4\,f^{2}}}}[{f_{,\zeta }}^{2}+({
\epsilon _{,\zeta }}-\psi\,{\chi _{,\zeta }})^2 
-f\,(\kappa ^{2}{\psi _{,\zeta }}^{2}+{\frac{1}{\kappa ^{2}}}\, {\chi _{,\zeta }}^{2})-({\frac{{2\,f}}{{
\alpha \,\kappa }}})^{2}\,{\kappa _{,\zeta }}^{2}], \label{eq:k}
\end{eqnarray}

Equations (\ref{eq:Pot}) are equivalent to equations (\ref{eq:Ein}). This new set of field equations can be obtained from the
Lagrangian 
\begin{equation}
{\cal L}={\frac{\rho }{{2\,f^{2}}}}\,[D\,f^{2}+(D\,\epsilon -\psi \,D\,\chi
)^{2}]-{\frac{{2\,\rho }}{{\alpha ^{2}\,\kappa ^{2}}}}\,D\,\kappa ^{2}-{
\frac{{\rho \,\kappa ^{2}}}{{2\,f}}}\,(D\psi ^{2}+{\frac{1}{{\kappa ^{4}}}}%
\,D\,\chi ^{2}).  \label{lag2}
\end{equation}
I have to do an important remark. The new Lagrangian (\ref{lag2}) cannot be obtained if the
transformation is made directly on the original Lagrangian, given by (\ref{lag0}) or (\ref{lag1}). This fact implies that the transformations
defined by (\ref{chi}), and (\ref{eps}) must have a degeneracy.

Lagrangian (\ref{lag2}) can be seen as obtained
from the line element $dS^{2}$ of a potential space, in other worlds, 
\begin{equation}
 {\cal L}\rightarrow dS^{2}=G_{AB}\,d\Psi ^{A}\,d\Psi ^{B},
\end{equation}
with $\Psi ^{A}=(f,\epsilon ,\chi
,\psi ,\kappa )$, so that the equations of motion (\ref{eq:Pot}),
obtained from variations of this Lagrangian with respect to the
coordinates, $\Psi ^{A}$, can be thought as geodesics in such
potential space 
\begin{equation}
dS^2={\frac{1}{{2\,f^{2}}}}\,[d\,f^{2}+(d\,\epsilon -\psi \,d\,\chi
)^{2}]-{\frac{{2}}{{\alpha ^{2}\,\kappa ^{2}}}}\,d\,\kappa ^{2}-{
\frac{{\kappa ^{2}}}{{2\,f}}}\,(d\psi ^{2}+{\frac{1}{{\kappa ^{4}}}}
\,d\,\chi ^{2}).  \label{dS2}
\end{equation}

As usual \cite{DoctorN}, metric (\ref{dS2}) defines a Riemannian potential space with constant scalar
curvature, $R=-12+\alpha ^{2}$. All the covariant derivatives of the
Riemann tensor are proportional to $\alpha ^{2}+3$, (or zero for $\alpha=0$). This fact is very important and let us decide which method we can use to solve the field equations. In what follows I give a method for solving the (\ref{dS2}) field equations with $\alpha\neq0$, where the formalism of the chiral equations can no be applied. 
The functional geodesic
formulation consist on defining an abstract space whose coordinates are defined by
the metric functions and the fields entering in the system. In order to
introduce an ansatz resembling the harmonic map ansatz into the functional
geodesic formulation (see \cite{kra}, and \cite{DHA} for an explanation) ,
we shortly explain the general idea of the harmonic map ansatz method. The
field equations of the theory can be written as 
\begin{equation}
D(\rho \,D\,\Psi^{A})+\rho \{_{B\;C}^{A}\}D\,\Psi^{B}\,D\,\Psi ^{C}=0  \label{diser}
\end{equation}
where $\Psi ^{A}$ are the potentials of the geodesic formulation and $\{_{B\;C}^{A}\}$ are the Christoffel symbols of the Riemannian space $dS^2$
defining the potential space of the theory. 

We can transform the field equations (\ref{eq:Pot}) into a set of first order differential equations defining the functions \cite{maridari}
\begin{eqnarray}
A &=&{\frac{{1}}{{2\,f}}}\,[D\,f-i\,(D\,\epsilon -\psi \,D\,\chi )],\nonumber \\
B &=&-{\frac{{1}}{{2\,\sqrt{f}}}}\,(\kappa \,D\,\psi -{\frac{i}{{\kappa }}}%
\,D\,\chi ),\nonumber \\
C &=&-{\frac{{D\,\kappa }}{{\kappa }}}\,
\label{eq:ABC}
\end{eqnarray}
so I can rewrite the equations (\ref{eq:Pot}) into the
following system: 
\begin{eqnarray}
{\frac{1}{{\rho }}}D(\rho \,A) &=&A\,(A-\bar A)+B\,\bar B,  \nonumber
\\
{\frac{1}{{\rho }}}D(\rho \,B) &=&{\frac{1}{{2}}}\,B\,(A-3\,\bar A)-C\,\bar B,  \nonumber
\\
{\frac{1}{{\rho }}}D(\rho \,C) &=&\frac{1}{2}\,\alpha ^{2}\,(B^{2}+{\bar B}^{2}),
\label{ABCeq}
\end{eqnarray}
where a bar over the functions denotes complex conjugate. Notice that in this way, I have
reduced the system of field equations to a set of three first order
differential equations for the three functions $A,B$, and $C$.

\section{The Hamiltonian}

In order to reformulate the
Einstein-Maxwell-Phantom field equations for arbitrary $\alpha\neq0 $ I cannot use the harmonic map
formulation, but I can try instead to mimic that formulation by means of
appropriately chosen ansatz. Following \cite{maridari} I perform a Legendre transformation and defining ``momenta'', as $P_{a}={\frac{{\partial {\cal L}}}{{\partial \,D\,\Psi ^{a}}}}$, I can construct a new function which is along the lines of the standard
Hamiltonian although in our case, it has not the usual properties of
evolution associated with the Hamiltonians. This Hamiltonian has the
explicit form: 
\begin{equation}
{\cal H}={\frac{{f^{2}}}{{2\,\rho }}}\,({P_{f}}^{2}+{P_{\epsilon }}^{2})-{
\frac{{\alpha ^{2}\,\kappa ^{2}}}{{8\,\rho }}}\,{P_{\kappa }}^{2}-{\frac{{f}}{{2\,\rho }}}\,\left[{\frac{{{P_{\psi }}^{2}}}{{\kappa ^{2}}}}+\kappa ^{2}\,({P_{\chi }}+\psi \,{P_{\epsilon }})^{2}\right] .  \label{ham}
\end{equation}

The equations of motion can be derived from $D\Psi ^{a}={\frac{{\partial {\cal H}}}{{%
\partial \,P_{a}}}}$, and $D\,P_{a}=-{\frac{{\partial {\cal H}}}{{\partial
\,\Psi ^{a}}}}$, I obtain 
\begin{eqnarray}
D\,f &=&{\frac{{f^{2}}}{{\rho }}}\,P_{f}, \nonumber\\
D\,\epsilon &=&{\frac{{f^{2}}}{{\rho }}}\,[P_{\epsilon }-{\frac{{\kappa
^{2}\,\psi }}{{f}}}\,({P_{\chi }}+\psi \,{P_{\epsilon }})],\nonumber \\
D\,\psi &=&-{\frac{{f}}{{\rho \,\kappa ^{2}}}}\,P_{\psi }, \nonumber\\
D\,\chi &=&-{\frac{{f\,\kappa ^{2}}}{{\rho }}}\,({P_{\chi }}+\psi \,{%
P_{\epsilon }}),\nonumber \\
D\,\kappa &=&-{\frac{{\alpha ^{2}\,\kappa ^{2}}}{{4\,\rho }}}\,P_{\kappa },
\end{eqnarray}
Analogously, for the momenta I obtain the following equations
\begin{eqnarray}
D\,P_{f} &=&-{\frac{{f}}{{\rho }}}\,({P_{f}}^{2}+{P_{\epsilon }}^{2})+{\frac{
1}{{2\,\rho }}}\,\left[ {\frac{{{P_{\psi }}^{2}}}{{\kappa ^{2}}}}+\kappa ^{2}\,({%
P_{\chi }}+\psi \,{P_{\epsilon }})^{2}\right] ,  \nonumber \\
D\,P_{\epsilon } &=&0,  \nonumber \\
D\,P_{\psi } &=&{\frac{{f\,\kappa ^{2}}}{{\rho }}}\,({P_{\chi }}+\psi \,{
P_{\epsilon }})\,P_{\epsilon },  \nonumber \\
D\,P_{\chi } &=&0,  \nonumber \\
D\,P_{\kappa } &=&-{\frac{{f}}{{\rho \,\kappa }}}\,\left[ {\frac{{{P_{\psi }}^{2}}
}{{\kappa ^{2}}}}-\kappa ^{2}\,({P_{\chi }}+\psi \,{P_{\epsilon }})^{2}\right]+{%
\frac{{\alpha ^{2}\,\kappa }}{{4\,\rho }}}\,{P_{\kappa }}^{2}.  \label{eqH}
\end{eqnarray}
In terms of the momenta, the functions (\ref{eq:ABC}) transform into
\begin{eqnarray}
A &=&{\frac{{f}}{{2\,\rho }}}\,(P_{f}-i\,P_{\epsilon }),\nonumber \\
B &=&{\frac{\sqrt{f}}{{2\,\rho }}}\,[{\frac{{P_{\psi }}}{{\kappa }}}
-i\,\kappa \,({P_{\chi }}+\psi \,{P_{\epsilon }})] , \nonumber\\
C &=&-{\frac{{\alpha ^{2}\,\kappa }}{{4\,\rho }}}\,P_{\kappa }
\end{eqnarray}

In what follows I apply the generalised harmonic map ansatz to the equations (\ref{ABCeq}) and give some classes of solutions.

\section{The Generalised Harmonic Maps Ansatz}
Now I look for invariant
transformations of the equations (\ref{diser}), $i.e.$, transformations of
the form $\Psi ^{A}=$ $\Psi ^{A}(\lambda ^{i})$ that leave the field
equations (\ref{diser}) invariant, where $\lambda ^{i}$ are potentials
fulfilling the same field equations (\ref{diser}). The potentials $\lambda
^{i}$ define the Riemannian space $V_{p}$. In terms of the potentials $%
\lambda ^{i}$, the field equations (\ref{diser}) read 
\begin{equation}
\rho \lbrack \Psi _{\,\ \,,i\,j}^{A}-\Gamma^ k_{i\,j}\,\Psi_{\,\,,k}^{A}+\{_{B\;C}^{A}\}\Psi _{\,\,,i}^{B}\Psi
_{\,\,,j}^{C}]D\,\lambda^{i}\,D\,\lambda^{j}+\Psi
_{\,\,,k}^{A}[D\,(\rho \,D\,\lambda ^{k})+\rho
\Gamma^ k_{i\,j}\,D\,\lambda^{i}\,D\,\lambda^j]=0,  \label{dieser2}
\end{equation}
where $,i={\partial /}\partial \lambda ^{i}$ and $\Gamma^ i_{j\,k}$ are the Christoffel symbols of $V_p$. In terms of the Christoffel
symbols of the abstract Riemannian space $V_{p}$, (\ref{dieser2}) reads 
\begin{eqnarray}
 \Psi _{\,\ \,,i;\,j}^{A}+\{_{B\;C}^{A}\}\Psi _{\,\,,i}^{B}\Psi
_{\,\,,j}^{C}&=&0  \label{dieser3}\\
D\,(\rho \,D\,\lambda ^{k})+\rho
\Gamma^ k_{i\,j}\,D\,\lambda^{i}\,D\,\lambda^j&=&0\label{dieser4}
\end{eqnarray}
where I have used the field equations for the $\lambda ^{i}$'s and the fact
that the $\lambda ^{i}$'s are linear independent. 

In terms of the complex variable $\varsigma =\rho +i\ z$ and ${\bar{\varsigma}}$ its complex
conjugated, equation (\ref{dieser4}) reads
\begin{equation}
 (\rho \lambda _{\,\,,\varsigma }^{k})_{,\bar{\varsigma}}+(\rho
\lambda _{\,\,,\bar{\varsigma}}^{k})_{,\varsigma }+2\rho\,\Gamma^ k_{i\,j}\,\lambda _{\,\,,\varsigma }^{i}\,\lambda _{\,\,,\bar{\varsigma}}^j=0
\end{equation}
The ansatz consist in to choose an apropiated $V_p$ potential space. Here I will study the $V_1$ and $V_2$ spaces because they are the most simple ones, but I will obtain solutions only for the $V_1$ subspaces.

We start with a two dimensional Riemannian spaces $V_{2}$ with
constant curvature, parametrising this Riemannian spaces with two harmonic
parameters $\lambda $, and $\tau $, such that $\lambda ,\tau \in %
\hbox{{I}\kern-.2em{\bf R}}$. The line element is 
\begin{equation}
ds^{2}={\frac{{2\,(d\,\lambda ^{2}\,+d\,\tau ^{2})}}{{(1-\sigma \,(\lambda
^{2}\,+\tau ^{2}))^{2}}}=}\frac{d\xi d\bar\xi}{(1-\sigma \xi \bar\xi)^{2}},  \label{le2d}
\end{equation}
where $\sigma $ is a real constant proportional to the potential space
curvature, and $\xi =\lambda +i\tau $, for the case of complex parameters.
We know that this is a maximally symmetric space, so it has three killing
vectors. If the electromagnetic field vanishes any value for $\alpha $ is
similar, because there is not interaction between scalar and electromagnetic
fields. But if
there is electromagnetic interaction the situation is different. As we
showed in \cite{maridari}, the subalgebras for the potential space with arbitrary $\alpha $,
with three Killing vectors, are such that one of them has to be set to zero,
so I conclude that if the electromagnetic field does not vanish the only
case of maximally symmetric $V_{2}$ that can be taken is the one with $%
\sigma =0$. In this case, the parameters satisfy the usual Laplace equation: 
$D(\rho \,D\,\lambda )=0,$ $D(\rho \,D\,\tau )=0.$ As in the harmonic maps
ansatz case, let us express the functions $A,B,C$ in terms of these
parameters as follows
\begin{eqnarray}
A &=&a_{1}(\lambda ,\tau )\,D\,\lambda +a_{2}(\lambda ,\tau )\,D\,\tau , 
\nonumber \\
B &=&b_{1}(\lambda ,\tau )\,D\,\lambda +b_{2}(\lambda ,\tau )\,D\,\tau , 
\nonumber \\
C &=&c_{1}(\lambda ,\tau )\,D\,\lambda +c_{2}(\lambda ,\tau )\,D\,\tau .
\label{ABC}
\end{eqnarray}
Using the harmonic equations, $i.e.$ the Laplace equation, for these
parameters in the field equations (\ref{ABCeq}), and recalling the fact that 
$(D\,\lambda )^{2},(D\,\tau )^{2}$, and $D\,\lambda \,D\,\tau $ are
independent functions, from the system of equations for $A,B,C$, (\ref{ABCeq}), I obtain the following set of equations: 
\begin{eqnarray}
{a_{1}}_{,\lambda }-a_{1}\,(a_{1}-{\bar a_{1}})-b_{1}\,{\bar b_{1}}\,
&=&0,  \nonumber \\
{b_{1}}_{,\lambda }-{\frac{{b_{1}}}{2}}\,(a_{1}-3\,{\bar a_{1}})+c_{1}\,{%
\bar b_{1}} &=&0,  \nonumber \\
{c_{1}}_{,\lambda }-\frac{1}{2}\,\alpha ^{2}\,({b_{1}}^{2}+{{\bar b_{1}}}^{2})\, &=&0,  \label{abc1}
\end{eqnarray}
\begin{eqnarray}
{a_{2}}_{,\tau }-a_{2}\,(a_{2}-{\bar a_{2}})-b_{2}\,{\bar b_{2}} &=&0,
\nonumber \\
{b_{2}}_{,\tau }-{\frac{{b_{2}}}{2}}\,(a_{2}-3\,{\bar a_{2}})+c_{2}\,{\bar b_{2}}\, &=&0,  \nonumber \\
{c_{2}}_{,\tau }-\frac{1}{2}\,\alpha ^{2}\,({b_{2}}^{2}+{{\bar b_{2}}}^{2})\, &=&0,  \label{abc2}
\end{eqnarray}
\begin{eqnarray}
{a_{1}}_{,\tau }+{a_{2}}_{,\lambda }-2\,a_{1}\,a_{2}+{\bar a_{1}}\,a_{2}+a_{1}\,{\bar a_{2}}-b_{1}\,{\bar b_{2}}-{\bar b_{1}}\,b_{2} &=&0,  \nonumber \\
2\,{b_{1}}_{,\tau }+2\,{b_{2}}_{,\lambda }+b_{2}\,(a_{1}-3\,{\bar a_{1}})+b_{1}\,(a_{2}-3\,{\bar a_{2}})+2\,c_{1}\,{\bar b_{2}}+2\,c_{2}\,{\bar b_{1}} &=&0,\newline
\nonumber \\
{c_{1}}_{,\tau }+{c_{2}}_{,\lambda }+\alpha ^{2}\,(b_{1}\,b_{2}+{\bar b_{1}}\,{\bar b_{2}}) &=&0.  \label{const}
\end{eqnarray}

Equations (\ref{abc1}, \ref{abc2}, \ref{const}) are equivalent to the field
equations (\ref{dieser2}). Taking the original potential also as functions
of the harmonic parameters, I can express the functions $A,B$, and $C$ from (\ref{ABC}) as
\begin{eqnarray}
A &=&{\frac{{1}}{{2\,f}}}\,[f_{,\lambda }-i\,(\epsilon _{,\lambda }-\psi
\,\chi _{,\lambda })]\,D\,\lambda +{\frac{{1}}{{2\,f}}}\,[f_{,\tau
}-i\,(\epsilon _{,\tau }-\psi \,\chi _{,\tau })]\,D\,\tau ,  \nonumber \\
B &=&-{\frac{{1}}{{2\,\sqrt{f}}}}\,(\kappa \,\psi _{,\lambda }-{\frac{i}{{%
\kappa }}}\,\chi _{,\lambda })\,D\,\lambda -{\frac{{1}}{{2\,\sqrt{f}}}}%
\,(\kappa \,\psi _{,\tau }-{\frac{i}{{\kappa }}}\,\chi _{,\tau })\,D\,\tau ,
\nonumber \\
C &=&{\frac{{\kappa _{,\lambda }}}{{\kappa }}}\,D\,\lambda +{\frac{{\kappa
_{,\tau }}}{{\kappa }}}\,D\,\tau ,  \label{ABC2}
\end{eqnarray}
from which, and from (\ref{ABC}), I can make the following
identification
\begin{eqnarray}
a_{1}={\frac{{1}}{{2\,f}}}\,[f_{,\lambda }-i\,(\epsilon _{,\lambda }-\psi
\,\chi _{,\lambda })];\ \ &&a_{2}={\frac{{1}}{{2\,f}}}\,[f_{,\tau
}-i\,(\epsilon _{,\tau }-\psi \,\chi _{,\tau })],  \nonumber \\
b_{1}=-{\frac{{1}}{{2\,\sqrt{f}}}}\,(\kappa \,\psi _{,\lambda }-{\frac{i}{{%
\kappa }}}\,\chi _{,\lambda }); &&\,\ \ b_{2}=-{\frac{{1}}{{2\,\sqrt{f}}}}%
\,(\kappa \,\psi _{,\tau }-{\frac{i}{{\kappa }}}\,\chi _{,\tau }),  \nonumber
\\
c_{1}=-{\frac{{\kappa _{,\lambda }}}{{\kappa }}};\,\ \ \ \ \ &&c_{2}=-{\frac{{%
\kappa _{,\tau }}}{{\kappa }}}.  \label{iden}
\end{eqnarray}
In what follows I proceed to present some solutions to the Einstein-Maxwell-Phantom
system in terms of the harmonic functions $\lambda $, $\tau $.

\section{One Dimensional Subspaces}

 In this work we will suppose that $a_{1},b_{1},c_{1}$ as functions of $\lambda $ only, and $a_{2},b_{2},c_{2}$ are zero. Equations (\ref{const}), being only constraint equations, are fulfilled with this functions, so I focus on the equations for the variables $a_{1},b_{1},c_{1}$. We obtain the following solutions.

\subsection{First Class of Solutions}

We start with most simple supposition, let the functions $a_{1}=a_0,b_{1}=b_0,c_{1}=c_0$ be real constants, then I obtain that if $a_0,c_0$ are arbitrary real numbers and $b_{0}=0$ this is a solution of the field equations. 
Using  equation (\ref{iden}), I obtain that the potentials are given by
\begin{eqnarray}
f &=&\,f_{0}e^{\frac{3}{4}r_{1}\lambda},  \nonumber \\
\kappa &=&\kappa _{0}\,e^{-\frac{1}{2}\alpha\,r_{2}\lambda},\nonumber\\
\epsilon &=&0,\,\,\,\,  \chi =0,\,\,\,\,\psi =0 
  \label{sol1}
\end{eqnarray}

Using the definition (\ref{eq:k}), (\ref{chi}) and (\ref{eps}) of the potentials, I can integrate them to the metric functions. We arrive at
\begin{eqnarray}
f &=&\,f_{0}e^{\frac{3}{4}\,r_1\,\lambda},\,\,\,\ \phi =r_2\,\lambda, \nonumber \\
k_{,\lambda}&=&\frac{\rho}{16\,\alpha^2}(r_2^2\,\alpha^2-9\,r_1^2)\nonumber\\
\omega &=&0,\,\,\,\,  A_0 =0,\,\,\,\,A_3 =0, 
  \label{sol1A}
\end{eqnarray}
that means, this solution contains no electromagnetic field and no 
rotation. For this solution, metric (\ref{papa_BL}) becomes
\begin{eqnarray}
ds^{2}&=&-f_{0}e^{\frac{3}{4}\,r_1\,\lambda}\,dt^{2}+K\,\frac{dr ^{2}}{f_{0}}e^{-\frac{3}{4}\,r_1\,\lambda}\nonumber\\
&+&(r^2-2\,m\,r+\sigma^2)\,\frac{1}{f_{0}}e^{-\frac{3}{4}\,r_1\,\lambda}\,\left(K\, d\theta^{2}+\sin^2(\theta)\,d\varphi ^{2}\right) ,  \label{papa_BL1}
\end{eqnarray}

One intersting example of this class is the following. We chose $\lambda_2$ from appendix \ref{App:Laplace} with $m_0=-\pi/2$,  that means
\begin{eqnarray}
\lambda&=&\ln\left(  \frac{r-m-\sqrt{\sigma^2-m^2}}{r-m+\sqrt{\sigma^2-m^2}}\right)-\frac{\pi}{2}\nonumber\\
&=&\tan^{-1}\left( \frac{r-m}{\sqrt{\sigma^2-m^2}}\right)-\frac{\pi}{2} 
\end{eqnarray}
and 
\[
  \frac{3}{4}\,r_1=-\phi_0\hskip1cm r_2=\sqrt{2+\frac{\phi_0^2}{2}}
\]
with $f_0=1$, metric (\ref{papa_BL1}) transform into
\begin{eqnarray}
ds^{2}&=&-e^{-\phi_0\,\lambda}\,dt^{2}+e^{\phi_0\,\lambda}\,\left[ dr ^{2}
+(r^2-2\,m\,r+\sigma^2)\,\left( d\theta^{2}+\sin^2(\theta)\,d\varphi ^{2}\right)\right]  ,  \nonumber\\
\phi&=&\sqrt{2+\frac{\phi_0^2}{2}}\,\lambda
\label{eq:sol_MN}
\end{eqnarray}
(remember that in this work I am using units such that $8\pi\,G/c^4=1$). Solution (\ref{eq:sol_MN}) is the static one given in \cite{dario}. This solution represents a pure scalar field in a static space-time. This solution contains the Morris-Thorn wormhole for $\phi_0=0,\,m=0$ (for more details see reference \cite{dario}).

\subsection{Second Class of Solutions}

We will suppose again that $a_{1}=a_0,b_{1}=b_0,c_{1}=c_0$ are constants, but now they can be complex numbers. We solve the field equations (\ref{abc1}) and obtain the solution 
\begin{eqnarray}
a  &=&\frac{1}{2}\,i{r_1},\nonumber\\ 
b&=& \frac{1}{2}\left( 1+\,i \right) {r_1},\nonumber\\
c  &=&-{r_1} 
\end{eqnarray}
Using equations (\ref{iden}) I arrive at
\begin{eqnarray}
  f &=&{f_0},\nonumber\\
\kappa  &=&{\kappa_0}\,{e^{{r_1}\,\lambda}},\nonumber\\
\psi  &=&-{\frac {{e^{-{\it r1}\,\lambda}}\sqrt {{f_0}}}{{\kappa_0}}}+{\psi_0},\nonumber\\
\chi &=&-\sqrt {{f_0}}{
\kappa_0}\,{e^{{r_1}\,\lambda}}+{\chi_0},\nonumber\\
\epsilon &=&-\sqrt {{f_0}}{e^{{r_1}\,\lambda}}{\psi_0}\,{
\kappa_0}+{\epsilon_0}
  \label{sol2}
\end{eqnarray}
where $f_0,\epsilon_0,\psi_0, \chi_0$ and $\kappa0$ are integration's constants. With these potential I can write down the corresponding components of the metric, I have
\begin{eqnarray}
f &=&{f_0},\,\,\,\, \phi=\frac{2\,r_1}{\alpha}\,\lambda\nonumber\\
k_{,\lambda}&=&-\frac{\rho\,r_1^2}{4\,\alpha^2}(\alpha^2+4)\nonumber\\
 A_0&=&-\frac{1}{2\,\kappa_0}\,\left( e^{-{r_1}\,\lambda}\sqrt { f_0
}-{\psi_0}\,{\kappa_0}\right) ,\nonumber\\
A_{3\,,{\lambda}}&=&-\frac{r_1}{2\,\sqrt {{f_0}}{\kappa_0}}
\, \left( \rho +\omega\, 
f_0 \right) e^{-{r_1}\,\lambda},\nonumber\\
\omega_{,{\lambda}}&=&-{\frac { {r_1}}{{
f_0}}} \rho 
  \label{sol2A}
\end{eqnarray}

Solution (\ref{sol2A}) is a rotating magnetised wormhole where the rotation generates an electric field. There is no gravitational potential, thus the space-time is curved by the charge and the Phantom field. We show an example, using equations (\ref{eqA:AOm}) I can integrate the functions $\omega$ and $A_3$. For example, I chose $\lambda_2$ from appendix \ref{App:Laplace}. The space time metric and the scalar field and electromagnetic potentials read
\begin{eqnarray}
 ds^{2}&=&-(dt-r_1\cos(\theta))^{2}\nonumber\\
&+&\left[ dr ^{2}
+(r^2-2\,m\,r+\sigma^2)\,\left( d\theta^{2}+\sin^2(\theta)\,d\varphi ^{2}\right)\right]  ,  \nonumber\\
\phi&=&\frac{2\,r_1}{\alpha}\,\lambda, \nonumber\\
A_0&=&-\frac{1}{2\,\kappa_0}\,\left( e^{-{r_1}\,\lambda}-{\psi_0}\,{\kappa_0}\right) ,\nonumber\\
A_3&=&-\cos \left( \theta \right) \frac{r_1}{\kappa_0}
\,{\exp\left( \frac{2\,{r_1}\,(\lambda+\frac{\pi}{2})}{\sqrt {
{\sigma}^{2}-{m}^{2}}} \right) }\nonumber\\
\lambda&=&\tan^{-1} \left( 
\frac {r-m}{\sqrt {{\sigma}^{2}-{m}^{2}}} \right)-\frac{\pi}{2}
\label{eq:sol_MorrisThorneMagn}
\end{eqnarray}
where I have set $f_0=1$ and $r_1^2(\alpha^2+4)=4\,\alpha^2$. In some seance, this is a generalisation of the Morris-Thorne solution, it is a magnetised and rotating wormhole. If I set $r_1=0$ metric (\ref{eq:sol_MorrisThorneMagn}) is flat. For $r\rightarrow\infty$ the magnetic field represents a magnetic monopole. If I want to generate magnetic dipoles in this solution, I have to use $\lambda_4$ or $\lambda_5$ from appendix \ref{App:Laplace} stead of $\lambda_2$.

\subsection{Third Class of Solutions}

We now suppose that one function, for example $b=0$ and solve the field equations (\ref{abc1}), I obtain the solution 
\begin{eqnarray}
a&=&\frac{-
\frac{1}{2}\left(l_1\,e^{2\,r_1\,\lambda}-l_2 \right)+i\,\sqrt{l_1\,l_2}\,e^{r_1\,\lambda}  }{l_1\,e^{2\,r_1\,\lambda}+l_2}
\, ,\nonumber\\ 
b&=& 0,\nonumber\\
c  &=&2\,{r_2} 
\end{eqnarray}
where $l_1,\,l_2,\,r_1$ and $r_2$ are integration's constants. Using equations (\ref{iden}) I arrive at
\begin{eqnarray}
  f &=&\frac{f_0\,e^{r_1\,\lambda}}{l_1\,e^{2\,r_1\,\lambda}+l_2},\nonumber\\
\kappa  &=&{\kappa_0}\,{e^{-2\,r_2\,\lambda}},\nonumber\\
\psi  &=&0\hskip0.5cm \chi =0,\nonumber\\
\epsilon &=&\frac{f_0\,\sqrt{\frac{l_1}{l_2}}}{l_1\,e^{2\,r_1\,\lambda}+l_2}
  \label{sol3}
\end{eqnarray}
where $f_0,\epsilon_0,\psi_0, \chi_0$ and $\kappa0$ are again integration's constants. With these potential I can write down the corresponding components of the metric, I have
\begin{eqnarray}
f &=&\frac{f_0\,e^{r_1\,\lambda}}{l_1\,e^{2\,r_1\,\lambda}+l_2},\nonumber\\
k_{,\lambda}&=&\left( \frac{r_1^2}{4}-\frac{4\,r_2^2}{\alpha^2}\right) \rho\,\nonumber\\
 A_0&=&0,\hskip0.5cm
A_{3}=0,\nonumber\\
\omega_{,{\lambda}}&=&-\frac {2\,\sqrt{l_1\,l_2}\,r_1}{f_0} \rho 
  \label{sol3A}
\end{eqnarray}

Solution (\ref{sol3A}) is a rotating phantom field. For this solution metric (\ref{papa_BL}) reads
\begin{eqnarray}
ds^{2}&=&-\frac{f_0\,e^{r_1\,\lambda}}{l_1\,e^{2\,r_1\,\lambda}+l_2}\,(dt-\omega\,d\varphi)^{2}\nonumber\\
&+&\frac{l_1\,e^{2\,r_1\,\lambda}+l_2}{f_0\,e^{r_1\,\lambda}}\,\left[ K\,dr ^{2}
+(r^2-2\,m\,r+\sigma^2)\left(K\, d\theta^{2}+\sin^2(\theta)\,d\varphi ^{2}\right)\right]  ,  \label{papa_BL32}
\end{eqnarray}

Again I chose $\lambda_2$ from appendix \ref{App:Laplace}, with $m_0=-\pi/2$ and 
\[
  \alpha^2=\frac{16\,r_2^2 }{1-r_1^2}\hskip1cm r_1=-\phi_0\hskip1cm r_2=\sqrt{2+\frac{\phi_0^2}{2}}
\]
with $f_0=1$, metric (\ref{papa_BL32}) transform into
\begin{eqnarray}
ds^{2}&=&-f\,(dt-2\,\sqrt{l_1\,l_2}\,\phi_0\,\cos(\theta)\,d\varphi)^{2}\nonumber\\
&+&\frac{1}{f}\,\left[ dr ^{2}
+(r^2-2\,m\,r+\sigma^2)\left(d\theta^{2}+\sin^2(\theta)\,d\varphi ^{2}\right)\right]   ,  \nonumber\\
f&=&\frac{e^{-\phi_0\,\lambda}}{l_1\,e^{-2\,\phi_0\,\lambda}+l_2}\hskip0.5cm\phi=\sqrt{2+\frac{\phi_0^2}{2}}\,\lambda
\label{eq:sol_MNrot}
\end{eqnarray}

Solution (\ref{eq:sol_MNrot}) is the rotating one given in \cite{dario}. This solution represents a pure wormhole in a rotating spherically symmetric space-time. This solution contains the static one by seating $l_1=0$ ($l_2=1$) in (\ref{eq:sol_MNrot}) (see ref. \cite{dario}). Solutions (\ref{eq:sol_MNrot}) is not asymptotically flat because of the rotation is provoked by the $\cos(\theta)$. This is because I chose the harmonic map $\lambda_3$. To avoid this problem I can choose for example the harmonic map $\lambda_5$ in appendix \ref{App:Laplace}. Then metric (\ref{eq:sol_MNrot}) becomes
\begin{eqnarray}
ds^{2}&=&-f\,(dt-\frac{2\,\sqrt{l_1\,l_2}\,\phi_0\,(r-m)\,\sin^2(\theta)}{\Delta}\,d\varphi)^{2}\nonumber\\
&+&\frac{1}{f}\,\left[ K\,dr ^{2}
+(r^2-2\,m\,r+\sigma^2)\left(K\,d\theta^{2}+\sin^2(\theta)\,d\varphi ^{2}\right)\right]   ,  \nonumber\\
f&=&\frac{e^{-\phi_0\,\lambda}}{l_1\,e^{-2\,\phi_0\,\lambda}+l_2}\hskip0.5cm\phi=\sqrt{2+\frac{\phi_0^2}{2}}\,\lambda\nonumber\\
 \lambda&=&\frac{\cos(\theta)}{\Delta}\hskip0.5cm K=\frac{\Delta}{r^2-2\,m\,r+\sigma^2}\nonumber\\
\Delta&=&(r-m)^2+(\sigma^2-m^2)\,\cos(\theta)
\label{eq:sol_MNrot2}
\end{eqnarray}
Metric (\ref{eq:sol_MNrot2}) is asymptotically flat if $l_1+l_2=1$. It can be shown that metric (\ref{eq:sol_MNrot2}) is an asymptotically flat rotating wormhole generated by phantom matter (see ref \cite{jmendez}).

\subsection{Forth Class of Solutions}

 Now I look for solutions of the form $a=(m_1\lambda+m_2)/(l_1\lambda+l_2))$ ... and in the same way for the other functions $b$ and $c$. We set these ansatz into equation (\ref{abc1}) and obtain the following solution
\begin{eqnarray}
 a &=& -\frac{1}{4}\,{\frac {3\,{
 r_1}\,{ l_1}\,\lambda+3\,{r_1}\,{l_2}+{l_1}}{{l_1
}\,\lambda+{l_2}}},\nonumber\\
b &=&\frac{1}{2}\,{\frac {i{\chi_0}
\,{l_1}}{\sqrt {{f_0}}\,{\kappa_0} \left( {l_1}\,\lambda+{l_2}
 \right) }}\nonumber\\
c &=&-\frac{3}{4}\,{\frac {{ 
r_1}\,{l_1}\,\lambda+{r_1}\,{l_2}-{l_1}}{{l_1}\,
\lambda+{l_2}}},
\end{eqnarray}
With these functions into (\ref{abc1}) I obtain the corresponding potentials given by
\begin{eqnarray}
f&=&{\frac {{f_0}\,{e^{-\frac{3}{2}\,
{r_1}\,\lambda}}}{\sqrt {{l_1}\,\lambda+{l_2}}}},\nonumber\\
\kappa &=&{\kappa_0}\, \left( {
\frac {{e^{{r_1}\,\lambda}}}{{l_1}\,\lambda+{l_2}}}
 \right) ^{3/4},\nonumber\\
\chi
 &=&{\frac {{\chi_0}}{{l_1}\,\lambda+{l_2
}}}\nonumber\\
 \psi &=&0,\,\,\,\,
\epsilon=0,
\label{sol3}
\end{eqnarray}
Using the definitions (\ref{iden}), I arrive to the metric components
\begin{eqnarray}
f&=&{\frac {{f_0}\,{e^{-\frac{3}{2}\,
{r_1}\,\lambda}}}{\sqrt {{l_1}\,\lambda+{l_2}}}},\nonumber\\
\phi &=&-\frac{3}{4\,\alpha}\left[ {r_1}\,\lambda-\ln\left( l_1\,\lambda+l_2\right)\right] \nonumber\\
k_{,\lambda}&=&\frac{3}{4}r_1^2\rho\nonumber\\
  A_{3\,,{\lambda}}&=&-\frac{1}{2}\,{\frac {  {\chi_0}\,{l_1}}{{f_0}\,{{\kappa_0}}^{2}}}\,\rho\nonumber\\
A_0&=&0,\,\,\,\,\ \omega=0.\label{sol3A}
\end{eqnarray}
Unfortunately, this solution requires $\alpha^2=-3$ and this implies that the scalar field is imaginary. Some explicit solutions of this class can be found in \cite{ivan}.

\subsection{Fifth Class of Solutions}

In this subsection I will introduce an ansatz not anymore in the functions $a,b,c$, but directly in the potentials. We start with the following ansatz
\begin{eqnarray}
f &=&{\frac {{f_0}}{
\sqrt {F}}},\hskip0.5cm
 \kappa ={\frac {{\kappa_0}}{F ^{3/4}}},\nonumber\\
\psi &=&{\psi_0},\hskip0.5cm
\chi ={\chi_0}\,{
\frac {F_{,\lambda}}{F}},\nonumber\\
\epsilon&=&{
\epsilon_0}\,{\frac {F_{,\lambda}}{F}} 
\label{sol4}
\end{eqnarray}
where $f_0,\kappa_0,\psi_0,\chi_0$ and $\epsilon_0$ are constants restricted to $\epsilon_0=\psi_0\,\chi_0$ and $f_0=\chi_0^2/\kappa_0^2$ and $F$ is a function of $\lambda$. Ansatz (\ref{sol4}) is a solution of the Einstein-Maxwell Phantom field equations if function $F$ fullfil the following differential equations
\begin{eqnarray}
 F_{,\lambda\,\lambda\,\lambda}\,F-F_{,\lambda\,\lambda}\,F_{,\lambda}=0,\nonumber\\
 \left( F_{,\lambda\,\lambda}\,F-F_{,\lambda}^2\right) \,\left( F_{,\lambda\,\lambda}\,F-F_{,\lambda}^2+1\right)=0
\nonumber\\
 \left( F_{,\lambda\,\lambda}\,F-F_{,\lambda}^2\right) \,\left( F_{,\lambda\,\lambda}\,F-F_{,\lambda}^2-\frac{3}{\alpha^2}\right)=0
\label{eq:F}
\end{eqnarray}
Fortunately the derivative of the equation $F_{,\lambda\,\lambda}\,F-F_{,\lambda}^2+F_0$ being $F_0$ an arbitrary constant, is the first equation in (\ref{eq:F}), therefore if the second and third equations of (\ref{eq:F}) are fullfiled, the first one do. Equations (\ref{eq:F}) have two solutions, $F_0=0$ and $\alpha$ arbitrary and $F_0=1$ and $\alpha^2=-3$. This differential equation can also be written as
\begin{equation}
 F^2\,\left(\frac{ F_{,\lambda}}{F}\right) _{,\lambda}=-F_0
\label{eq:F0}
\end{equation}
Some solutions of equations (\ref{eq:F0}) are 
\begin{eqnarray}
 F&=&\sqrt{F_0}\,\lambda+l_2\nonumber\\
F&=&\frac{1}{\Lambda}\,\sqrt{\frac{F_0}{2}}\,(\sin(\Lambda\,\lambda)+\cos(\Lambda\,\lambda))\nonumber\\
F&=&-\frac{1}{4\,\Lambda}\,e^{\Lambda\,\lambda}+F_0\,e^{-\Lambda\,\lambda}
\end{eqnarray}
 
With this solution the space-time metric componets read
\begin{eqnarray}
f &=&{\frac {{f_0}}{
\sqrt {F}}},\hskip0.5cm
 \phi =\frac{3}{4\,\alpha}\ln(F) ,\nonumber\\
A0&=&\frac{1}{2}\,{\psi_0},\hskip0.5cm
 A_{3\,,\lambda}=-\frac{1}{2}\,\frac {F_0 }{\chi_0}\,\rho ,\nonumber\\
k&=&0,\hskip0.5cm\omega=0 
\label{sol4A}
\end{eqnarray}

Unfortunatly the interesting solutions with $\alpha$ arbitrary implies $F_0=0$. The solution $F_0=1$ implies again that the phantom field is imaginary. With this step the space-time metric is complitely integrated. Solution (\ref{sol4A}) represets a static, magnetized Phantom field. The space-time metric for this case reads
\begin{eqnarray}
 ds^{2}&=&-f\,dt^{2}
+\frac{1}{f}\,\left[K\, dr ^{2}
+(r^2-2\,m\,r+\sigma^2)\left(K\,d\theta^{2}+\sin^2(\theta)\,d\varphi ^{2}\right)\right]   ,  \nonumber\\
f &=&{\frac {{f_0}}{
\sqrt {F}}},\hskip0.5cm\phi=\frac{3}{4\,\alpha}\ln(F) ,\nonumber\\
 K&=&\frac{(r-m)^2+(\sigma^2-m^2)\,\cos^2(\theta)}{(r^2-2\,m\,r+\sigma^2)}
\label{eq:sol_Magn}
\end{eqnarray}

Let us see one example, let be  $F=-\frac{1}{4\,\Lambda}\,e^{\Lambda\,\lambda}+F_0\,e^{-\Lambda\,\lambda}$. Then components of the space-time metric reads
\begin{eqnarray}
f &=&{\frac {{f_0}}{
\sqrt {-\frac{1}{4\,\Lambda}\,e^{\Lambda\,\lambda}+F_0\,e^{-\Lambda\,\lambda}}}},\hskip0.5cm\phi=\frac{3}{4\,\alpha}\ln(-\frac{1}{4\,\Lambda}\,e^{\Lambda\,\lambda}+F_0\,e^{-\Lambda\,\lambda}) ,\nonumber\\
 K&=&\frac{(r-m)^2+(\sigma^2-m^2)\,\cos^2(\theta)}{(r^2-2\,m\,r+\sigma^2)}
\label{eq:sol_Magn}
\end{eqnarray}
If I choose $\lambda_1$ from appendix \ref{App:Laplace}, the correspondig magnetic field reads
\begin{equation}
 A_3=F_0\,\frac{m}{\chi_0}\,(\cos(\theta)-1),
\end{equation}
which corresponds to a magnetic monopol. If I want the magnetic field of a magnetic dipole stead of the magnetic monopole I have to use $\lambda_4$ or $\lambda_5$ from appendix \ref{App:Laplace}.

\section{Conclusions}

In this work I have given a method from which we can derive a set of formulas which can be integrated in order to fiend exact solutions of the Einstein-Maxell Phantom fields. We have generated rotating and magnetised solutions, in particular we gave the rotating and magnetised generalisations of the Morris-Thorne solution \cite{MT}. But we have given some other solutions with similar behaviour as other ones (see for example \cite{topics}). In order to find new exact solutions, we have to follow the receipt: 

1.- Choose a class in terms of $\lambda$ with the features you need.

2.- Choose an harmonic map from appendix \ref{App:Laplace} with the features you need.

3.- From appendix \ref{App:Laplace} you have also the integration of the rotation and most of the time of the magnetic field. 

Finally write the space-time metric and analyse the solution. With this method we can further generate magnetic monopoles, quadrupoles, etc. cupled with gravitational multipoles. This give rise to complete the study of the wormholes and see if there exist some stables ones. 

\section*{acknowledgements} I would like to thank
Dar\'{\i}o N\'u\~nez, Ivan Rodr\'{\i}guez and Juan M\'endez for
many helpful and useful discussions. The numerical computations were
carried out in the "Laboratorio de Super-C\'omputo
Astrof\'{\i}sico (LaSumA) del Cinvestav" and in the UNAM's cluster Kan-Balam. This work was partly
supported by CONACyT M\'exico, under grants 49865-F, 54576-F,
56159-F, and by grant number I0101/131/07 C-234/07, Instituto
Avanzado de Cosmologia (IAC) collaboration.

\appendix

\section[]{The Laplace Equation}\label{App:Laplace}

In this appendix I give some generalities of the Lapace equation

\begin{equation}
 D(\rho D\lambda)=0
\end{equation}
In coordinates $(\rho, z)$, this equation reads
\begin{equation}
 \rho\,( \lambda_{,\rho\rho}+\frac{1}{\rho}\lambda_{,\rho}+\lambda_{,zz})=0
\end{equation}
or in coordinates $\zeta=\rho+i\,z$, it is
\begin{equation}
 (\rho\lambda_{,\zeta})_{,\bar\zeta}+(\rho\lambda_{,\bar\zeta})_{,\zeta}=0
\end{equation}
Finally, in Boyer-Lindquist coordinates
 $\rho=\sqrt{r^2-2\,m\,r+\sigma^2}\,\sin(\theta), z=(r-m)\,\cos(\theta)$, the Laplace equations is given by
\begin{equation}
 ((r^2-2\,m\,r+\sigma^2)\lambda_{,r})_{,r}+\frac{1}{\sin(\theta)}(\sin(\theta)\lambda_{,\theta})_{,\theta}=0
\end{equation}

where $m$ and $\sigma$ are arbitrary constans. Some exact solutions of the Laplace equation in the previous coordinates are the following.
\begin{eqnarray}
\lambda_1&=&\lambda_0\,\ln\left(1-\frac{2\,m}{r} \right)+m_0, \hskip1cm\sigma=0  \nonumber\\
\lambda_2&=&\lambda_0\,\tan^{-1}\left( \frac{r-m}{\sqrt{\sigma^2-m^2}}\right)+m_0 \hskip1cm\sigma^2\geqslant m^2\nonumber\\
\lambda_3&=&\lambda_0\,\ln\left(  \frac{r-m-\sqrt{m^2-\sigma^2}}{r-m+\sqrt{m^2-\sigma^2}}\right)+m_0 \hskip1cm\sigma^2\leqslant m^2\hskip0.5cm\text{or}\nonumber\\
\lambda_3&=&\lambda_0\,\tanh^{-1}\left( \frac{r-m}{\sqrt{m^2-\sigma^2}}\right)+m_0 \hskip1cm\sigma^2\leqslant m^2\nonumber\\
\lambda_4&=&\lambda_0\,\frac{r-m}{\Delta}+m_0\hskip1cm\Delta=(r-m)^2+(\sigma^2-m^2)\,\cos(\theta)\nonumber\\
\lambda_5&=&\lambda_0\,\frac{\cos(\theta)}{\Delta}+m_0\nonumber\\
\lambda_6&=&\lambda_0\,\ln\left((r^2-2\,m\,r+\sigma^2)\,\sin^2(\theta) \right)+m_0 \nonumber\\
\lambda_7&=&\lambda_0\,\ln\left(\tan\left( \frac{1}{2}\,\theta\right) \right) +m_0
\end{eqnarray}
where $\lambda_0$ and $m_0$ are also arbitrary constants. In the text I have functions $A=A(\rho,z)=A(r,\theta)$, or using the harmonic map, this equivalent to $A=A(\lambda(r,\theta))$, such that the derivative is $A_{,r}=A_{,\lambda}\,\lambda_{,r}$. The field equation for $\omega$ is
$\omega_{,\zeta}=(\epsilon_{,\lambda}-\psi\,\chi_{,\lambda})\,\lambda_{,\bar\zeta}=A_{,\lambda}\,\lambda_{,\bar\zeta}$ and for $A_{3,\zeta}=\frac{\rho}{2\,f\,\kappa^2}\chi_{,\bar\zeta}+\omega\,\psi_{,\zeta}$. In terms of the Boyer-Lindquist  coordinates $(r,\theta)$, the  differential equation for the rotations component $\omega$ and the magnetic potential $A_3$ is
\begin{eqnarray}
\omega_{,r}&=&-(\epsilon_{,\lambda}-\psi\,\chi_{,\lambda})\,\sin(\theta)\,\lambda_{,\theta}\nonumber\\
\omega_{,\theta}&=&(\epsilon_{,\lambda}-\psi\,\chi_{,\lambda})\,\,(r^2-2\,m\,r+\sigma^2)\,\sin(\theta)\,\lambda_{,r}\\
A_{3,r}&=&-\frac{\rho}{2\,f\,\kappa^2}\chi_{,\lambda}\,\sin(\theta)\,\lambda_{,\theta}+\omega\,\psi_{,\lambda}\,\lambda_{,r}\nonumber\\
A_{3,\theta}&=&\frac{\rho}{2\,f\,\kappa^2}\chi_{,\lambda}\,(r^2-2\,m\,r+\sigma^2)\,\sin(\theta)\,\lambda_{,r}+\omega\,\psi_{,\lambda}\,\lambda_{,\theta}
\label{eqA:AOm}
\end{eqnarray}
Generically I have to integrate a differential equations of the form
\begin{eqnarray}
A_{,r}&=&-A_0\,\sin(\theta)\,\lambda_{,\theta}\nonumber\\
A_{,\theta}&=&A_0\,(r^2-2\,m\,r+\sigma^2)\,\sin(\theta)\,\lambda_{,r}
\label{eqA:A}
\end{eqnarray}
where $\rho\,A_0=A_{,\lambda}$, and $A$ here stands for $A=A_3,\omega$, respectively. We can integrate the (\ref{eqA:A}) for each $\lambda$, I arrive at
\begin{eqnarray}
\lambda_1 \Rightarrow
A&=&-2\,\lambda_0\,A_0\,m\,\cos(\theta)-\lambda_1\nonumber\\
\lambda_2 \Rightarrow
A&=&-\lambda_0\,A_0\,\sqrt{\sigma^2-m^2}\,\cos(\theta)-\lambda_1\nonumber\\
\lambda_3 \Rightarrow A&=&-\lambda_0\,A_0\,\sqrt{m^2-\sigma^2}\,\cos(\theta)-\lambda_1\nonumber\\
\lambda_4 \Rightarrow A&=&\lambda_0\,A_0\,\left( \frac{(\sigma^2-m^2)\sin^2(\theta)\,\cos(\theta)}{\Delta}+\cos(\theta)-\lambda_1\right) \nonumber\\
\lambda_5 \Rightarrow A&=&\lambda_0\,A_0\,\frac{(r-m)\,\sin^2(\theta)}{\Delta}-\lambda_1\nonumber\\
\lambda_6 \Rightarrow A&=&-2\,\lambda_0\,A_0\,(r-m)\,\cos(\theta)-\lambda_1\nonumber\\
\lambda_7 \Rightarrow A&=&\lambda_0\,A_0\,(r-m)-\lambda_1
\end{eqnarray}

 The other function to integrate is $k$. The differential equation for $k$ is
\begin{equation}
 k_{,\zeta}=k_{,\lambda}\,{\lambda_{,\zeta}}^2
\end{equation}
In term of the Boyer-Lindquist coordinates this differential equation reads
\begin{eqnarray}
 k_{,\theta} &=&{k_0}{\frac { \left( {r}^{2}-2\,mr+{\sigma}^{2} \right) \sin \left( \theta \right) \, \left(  -\left(  \left( {r}^{2}-2\,mr+{\sigma}^{2} \right)  \lambda_{,r} ^{2}- \lambda_{,\theta} ^{2} \right) \cos \left( \theta \right) +  2\,\left(r-m \right)  \lambda_{,\theta}\lambda_{,r} \sin \left( \theta \right)  \right) }{\left( r-m \right) ^{2}+\left( \sigma^2-m^2 \right) 
\cos ^{2}\left( \theta \right) }},\nonumber\\
k_{,r} &=&{
k_0}{\frac {\,\sin \left( \theta \right)  \left(  2\,\left( {r}^{2}-2\,mr+
{\sigma}^{2} \right)  \lambda_{,\theta} \lambda_{,r} \cos \left( 
\theta \right) +  \left( r-m \right) \left(  \left( {r}^{2}-2\,mr+{
\sigma}^{2} \right) \lambda_{,r}^{2}- \lambda_{,\theta}^{2} \right) \sin \left( \theta \right)  \right) }{ \left( r-m \right) ^{2}+\left( \sigma^2-m^2 \right) 
\cos ^{2}\left( \theta \right)}} 
\end{eqnarray}

We can integrate this differential equation for different $\lambda$'s. We obtain
\begin{eqnarray}
\lambda_1 \Rightarrow
k&=&2\,{k_0}\,{{\lambda_0}}^{2}\ln 
 \left( {\frac { {r}^{2}-2\,mr  }{ \left( r-m
 \right) ^{2}-{m}^{2}\cos^{2} \left( \theta \right) }
} \right) 
\nonumber\\
K&=& \left( {\frac { {r}^{2}-2\,mr  }{ \left( r-m
 \right) ^{2}-{m}^{2}\cos^{2} \left( \theta \right) }
} \right)^{4\,{k_0}\,{\lambda_0}^{2}-1 }
 \nonumber\\
\lambda_2 \Rightarrow
k&=&-\frac{1}{2}\,{k_0}\,{{\lambda_0}}^{2}\ln 
 \left( {\frac { {r}^{2}-2\,mr+{\sigma}^{2} }{
 \left( r-m \right) ^{2}+ \left( {\sigma}^{2}-{m}^{2} \right) 
\cos^{2} \left( \theta \right)  }} \right) \nonumber\\
K&=&\left( {\frac { {r}^{2}-2\,mr+{\sigma}^{2} }{
 \left( r-m \right) ^{2}+ \left( {\sigma}^{2}-{m}^{2} \right) 
\cos^{2} \left( \theta \right)  }} \right)^{-\,{k_0}\,{{\lambda_0}}^{2}-1}\nonumber\\
\lambda_3 \Rightarrow k&=&2\,{k_0}\,{{\lambda_0}}^{2}\ln 
 \left( {\frac {  {r}^{2}-2\,mr+{\sigma}^{2}}{
 \left( r-m \right) ^{2}+ \left( {\sigma}^{2}-{m}^{2} \right) 
\cos ^{2} \left( \theta \right)}} \right)\nonumber\\
K&=& \left( {\frac {  {r}^{2}-2\,mr+{\sigma}^{2}}{
 \left( r-m \right) ^{2}+ \left( {\sigma}^{2}-{m}^{2} \right) 
\cos ^{2} \left( \theta \right)}} \right)^{4\,{k_0}\,{{\lambda_0}}^{2}-1}\nonumber\\
\lambda_6 \Rightarrow 
k&=&2\,{{\lambda_0}}^{2}{k_0}\,\ln 
 \left(  \left( {r}^{2}-2\,mr+{\sigma}^{2} \right)  \sin^{2}
 \left( \theta \right)  \right) \nonumber\\
K&=&\frac{\left( r-m \right) ^{2}+ \left( {\sigma}^{2}-{m}^{2} \right) 
\cos ^{2} \left( \theta \right)}{{r}^{2}-2\,mr+{\sigma}^{2}}\left(  \left( {r}^{2}-2\,mr+{\sigma}^{2} \right)  \sin^{2}
 \left( \theta \right)  \right)^{4\,{{\lambda_0}}^{2}{k_0}}\nonumber\\
\lambda_7 \Rightarrow k&=&\frac{1}{2}\,{{\lambda_0}}^{2}{k_0}\,\ln 
 \left( {\frac { \sin  ^{2}\left( \theta \right)}
{ \left( r-m \right) ^{2}+ \left( {\sigma}^{2}-{m}^{2} \right) 
  \cos ^{2} \left( \theta \right)}} \right)\nonumber\\
K&=&\frac{\left( \sin \left( \theta \right)\right)^{2\,{\lambda_0}^{2}{k_0}}}{{r}^{2}-2\,mr+{\sigma}^{2}}  \left( {\frac { 1}
{ \left( r-m \right) ^{2}+ \left( {\sigma}^{2}-{m}^{2} \right) 
  \cos ^{2} \left( \theta \right)}} \right)^{{\lambda_0}^{2}{k_0}-1}
\end{eqnarray}
where 
\[
 K=\frac{(r-m)^2+(\sigma^2-m^2)\,\cos^2(\theta)}{(r^2-2\,m\,r+\sigma^2)}\,e^{2k}
\]

One use to choose $k_0$ in order to obtain $K=1$ or some simple expression.

\end{document}